\documentclass[useAMS,usenatbib]{mn2e}
\usepackage{epsfig,amsmath,amssymb}
\usepackage{graphicx}
\newcommand{\be}{\begin{equation}}    
\newcommand{\ee}{\end{equation}}
\newcommand{\beq}{\begin{eqnarray}}
\newcommand{\eeq}{\end{eqnarray}}
\newcommand{\beqn}{\begin{eqnarray*}}
\newcommand{\eeqn}{\end{eqnarray*}}

\title[Gravitational wave backgrounds and the cosmic transition from Population III to
Population II stars]{Gravitational wave backgrounds and the cosmic transition from Population III to Population II stars}

\author[Stefania Marassi, Raffaella Schneider, Valeria Ferrari]{Stefania Marassi$^{1}$\thanks{E-mail:
stefania.marassi@roma1.infn.it}, Raffaella Schneider$^{2}$, Valeria Ferrari$^{1}$\\
$^{1}$ Dipartimento di Fisica 'G. Marconi', Sapienza Universit\`a di
Roma,  Piazzale Aldo Moro 5, Roma, 00185, Italy\\
$^{2}$Osservatorio Astrofisico di Arcetri, Largo Enrico Fermi 5, 50125
Firenze, Italy}

\begin{document}

\date{Accepted 2009 . Received 2009 ; in original form 2009}

\pagerange{\pageref{firstpage}--\pageref{lastpage}} \pubyear{2009}

\maketitle

\label{firstpage}

\begin{abstract}
Using the results of a numerical simulation which follows the evolution, 
metal enrichment and energy deposition of both Population~III and Population~II stars,
we predict the redshift dependence of the formation rate of black hole remnants of 
Population~III stars with masses $100 - 500 M_{\sun}$ and of neutron stars (black
holes) remnants of Population II stars with masses $8 - 20 M_{\sun}$ ($20 - 40 M_{\sun}$).

We describe the gravitational wave spectrum produced by Population~III and Population~II 
sources adopting the most appropriate signals available in the literature and we compute 
the stochastic backgrounds resulting from the cumulative emission of these sources throughout
the history of the Universe.     

With the aim of assessing whether these backgrounds might act as foregrounds for signals 
generated in the Inflationary epoch, we compare their amplitudes with the sensitivity of currently
planned and future ground/space-based interferometers.

The predicted Population~III background lies in the sensitivity range of Ultimate-DECIGO, 
adding as a confusion-limited noise, with a peak amplitude of $\Omega_{\rm GW} h^2\simeq 3\times 
10^{-15}$ at $f=2.74$~Hz. However, differently to previous claims, we find that the gravitational wave 
background generated in the Inflationary epoch may dominate for $f\leq 2$ Hz. 
At frequencies $f \geq 10$~Hz, the background generated by Population~II stellar
remnants is much larger than that associated with Population~III stars, with peak 
amplitudes ranging between $10^{-12} \leq \Omega_{\rm GW} h^2 \leq 7\times 10^{-10}$ at frequencies
$f\in (387-850)$~Hz: progenitors with masses in the range $20 - 100 M_{\sun}$ contribute with a 
nearly monotonically increasing behavior, whereas for stars with masses $8 - 25 M_{\sun}$ 
the resulting background shape depends on the waveforms used to represent the collapse signal.  

Finally, we explore a scenario in which Super Massive Stars, formed 
in protogalactic dark matter halos out of gas of primordial composition and 
irradiated by a strong ultraviolet background, collapse to Super Massive 
Black Holes (SMBHs). Even assuming, as an upper limit to their formation rate, that
the mass density of these SMBHs at $z=10$ equals the presently observed value, 
the resulting gravitational wave background is too low to be detected 
from the space-based interferometer LISA; it is in the sensitivity range of 
Ultimate-DECIGO but in a frequency region seriously limited by the 
Galactic binary confusion background. 

\end{abstract}


\section{Introduction}

Gravitational waves of cosmological origin could be the result of a
large variety of astrophysical and cosmological processes that
develop in the very early Universe. As a consequence, 
our high redshift Universe is expected to be permeated with a background 
of gravitational radiation.
Depending on their origin, these stochastic gravitational wave
backgrounds (GWBs) will show different spectral properties and 
features that it is important to investigate in view of 
a possible, future detection.
 
At present, several gravitational waves interferometers 
(VIRGO, LIGO, GEO600, TAMA) are operating and taking data 
in the frequency range $\approx 10~{\rm Hz}-3~{\rm kHz}$.
In the near future, the low-frequency window will be accessible to observations: 
the Laser Interferometer Space Antenna\footnote{http://lisa.nasa.gov/} 
(LISA) covering the frequency range (0.1-100)mHz will be launched, and 
next-generation space detectors targeting (0.1-1)Hz 
as Decihertz Interferometer Gravitational wave Observatory (DECIGO)
and Big Bang Observer\footnote{http://universe.nasa.gov/program/vision.html} 
(BBO) are being planned \citep{Seto,KTHH}. 

These experiments will have as possible objectives to detect
or to set bounds on {\it primordial} GWBs, originating at 
the Big Bang, as well as backgrounds produced by the cumulative 
emission of {\it astrophysical} sources throughout the history 
of the Universe. In particular, the low-frequency region is where we expect the
contribution of the GWB from Population~III stars (Pop~III) that 
are the first stars to form in the Universe.
Pop~III stars do not contain metals and their formation and evolution 
may be different from that of stars of later generations (see
\citealt{BL} and \citealt{G} for thorough reviews of the subject). 
In particular, recent theoretical studies suggest that below a 
critical metallicity of $Z_{\rm cr} = 10^{-5 \pm 1} Z_{\sun}$
the reduced cooling efficiency and large accretion rates favor the formation
of massive stars, with characteristic masses $> 100 M_{\sun}$ 
(\citealt{O}; \citealt{BFCL}; \citealt{SFNO,SFSOB,SOIF}; \citealt{OTSF}). 

These stars are predicted to collapse to 
black holes of comparable masses (except in the mass range
$140 - 260 M_{\sun}$ where they are completely disrupted in pair-instability
supernova explosions, \citealt{HW}) and therefore are expected to
be efficient sources of gravitational waves.     
   
An earlier estimate of the GWB produced by these stars was done by \citet{SFCFM}
and \citet{AMA1}, in addition to that associated to core-collapse 
supernovae \citep{FMS1, FMS2, AMA2}. However, these earlier studies were limited
by our poor understanding of the characteristic masses and formation rates of Pop~III
stars. More recently, \citet{BSRJM}, \citet{SODV} and \citet{STKS1} 
pointed out that in some regions of the parameter-space and depending 
on the adopted Pop~III star formation rate, the GWB produced by the collapse 
of these first stars could be comparable or mask almost completely 
the primordial background predicted by standard Inflationary models.  
In addition, \citet{BSRJM} and \citet{SODV} have taken 
into account the contribution associated to core-collapse supernova events 
(the end-product of Pop~II stars with progenitor masses $> 8 M_{\sun}$) 
in order to investigate in which frequency ranges these two components could 
be discriminated. 

In order to give a realistic estimate of the GWB spectrum produced by the formation 
of the black holes remnants of the first stars, it is necessary to (i) reduce the big 
uncertainty related to the prediction of Pop~III star formation rate and (ii) adopt 
suitable models for the gravitational wave signals emitted during the collapse of these 
primordial stars. 

This paper is an attempt to satisfy these two requirements.
We compute the GWB produced by remnants of Pop~III stars in the mass range (100-500)$M_{\sun}$
using the Pop~III cosmic star formation rate obtained by \citet{TFS} in a recent numerical
simulation which follows the evolution, metal enrichment and energy 
deposition of both Pop~III 
and Pop~II stars, and give a reliable estimate of their relative rates at high redshift. 
As will be discussed later, we adopt as a template for the gravitational
waveforms associated to the collapse of Pop~III stars the results of recent numerical simulations
by \citet{STKS1} and discuss the related uncertainties comparing with other independent studies 
\citep{FWH,LSS,STKS2}. 
We also evaluate the contribution to the GWB of Pop~II stars, which explode as
core-collapse supernovae leaving a neutron star (progenitor masses in the range 
$[8-20]M_{\sun}$) or a black hole remnant (progenitor masses between $[20-40]M_{\sun}$), or
which directly collapse to black holes (progenitor masses larger than $40 M_{\sun}$).
In these ranges of Pop~II progenitor masses we choose the most
appropriate gravitational wave signals available in the literature.

The plan of the paper is as follows. In Section \ref{sectionsfr} 
we briefly describe the numerical simulation performed by \citet{TFS} 
and the model for the Pop~III/Pop~II cosmic star formation rate evolution 
that we have adopted in our analysis. In Sections \ref{sectionGWPopIII} and 
\ref{sectionGWPopII} we sketch out the main features of the waveforms which
describe the gravitational emission of single Pop~III and Pop~II sources. 
In Section \ref{sectionGWB} we present the density parameter, $\Omega_{\rm GW}$, 
of the GWBs, discuss their detectability by existing and planned detectors and 
the possibility that they might be foregrounds limiting the detection of 
primordial GWBs. In Section \ref{sectionSMS} we explore the GWB
produced by Super Massive Star formed in metal-free protogalaxies at
redshift $z=10$. Finally, in Section \ref{sectionconclusion} 
we draw our conclusions.

Throughout our work we have adopted a $\Lambda$CDM cosmological model
with parameters $\Omega_M=0.26$, $\Omega_\Lambda=0.74$, $h=0.73$, $\Omega_b=0.041$, 
in agreement with the three-year WMAP results \citep{S}. 

\section[]{The Population III/Population II cosmic star formation rate}
\label{sectionsfr}

Our present understanding of Pop~III star formation suggests that these
stars are not necessarily confined to form in the first dark
matter halos at redshift $z > 20$ but may continue to form during
cosmic evolution in regions of sufficiently low metallicity, with 
$Z < Z_{\rm cr}$. Thus, the Pop~III star formation rate and the
cosmic transition between Pop~III and Pop~II stars is regulated by the rate
at which metals are formed and mixed in the gas surrounding the first 
star forming regions, a mechanism that we generally refer to as chemical feedback.

Semi-analytic studies which implement chemical feedback generally find that, due
to inhomogeneous metal enrichment, the transition is extended in time, with coeval 
epochs of Pop~III and Pop~II star formation, and that Pop~III stars can continue to
form down to moderate redshifts, $z < 5$ (\citealt{SSF}; 
\citealt{FL}; \citealt{SSFC}). Yet, a direct observational 
evidence for the existence of Pop~III stars is still lacking. 

To better assess the validity of these semi-analytic models, \citet{TFS}
have performed a set of cosmological hydrodynamic simulations with an improved 
treatment of chemical enrichment \citep{TBDM}. In the simulation, it is
possible to assign a metallicity-dependent stellar initial mass function (IMF). When
$Z>Z_{\rm cr}$ Pop~II stars are assumed to form according to a Salpeter IMF 
$\Phi(M)\propto M^{-(1+x)}$ with $x=1.35$ and lower 
(upper) mass limit of $0.1 M_{\sun}$ ($100 M_{\sun}$). Stars with masses in the $8-40 M_{\sun}$
range explode as Type-II SNe and the explosion energy and metallicity-dependent metal yields 
are taken from \citet{WW}. Stars with masses $>40 M_{\sun}$ do not contribute
to metal enrichment as they are assumed to directly collapse to black holes. 
Very massive Pop~III stars form in regions where $Z < Z_{\rm cr}$; since theoretical models 
do not yet provide any indication on the shape of the Pop~III IMF, \citet{TFS} 
adopt a Salpeter IMF shifted to the mass range $100 - 500 M_{\sun}$; only stars in the 
pair-instability range ($140 - 260 M_{\sun}$) contribute to metal-enrichment and the
metal yields and explosion energies are taken from \citet{HW}. 
The simulation allows to follow metal enrichment properly accounting for the finite
stellar lifetimes of stars of different masses, the change of the stellar IMF and
metal yields. The numerical schemes adopted to simulate metal transport ad diffusion
are discussed in \citet{TBDM} to which we refer the interested reader 
for more details.  

\begin{figure}
\includegraphics[width=8.5cm,angle=360]{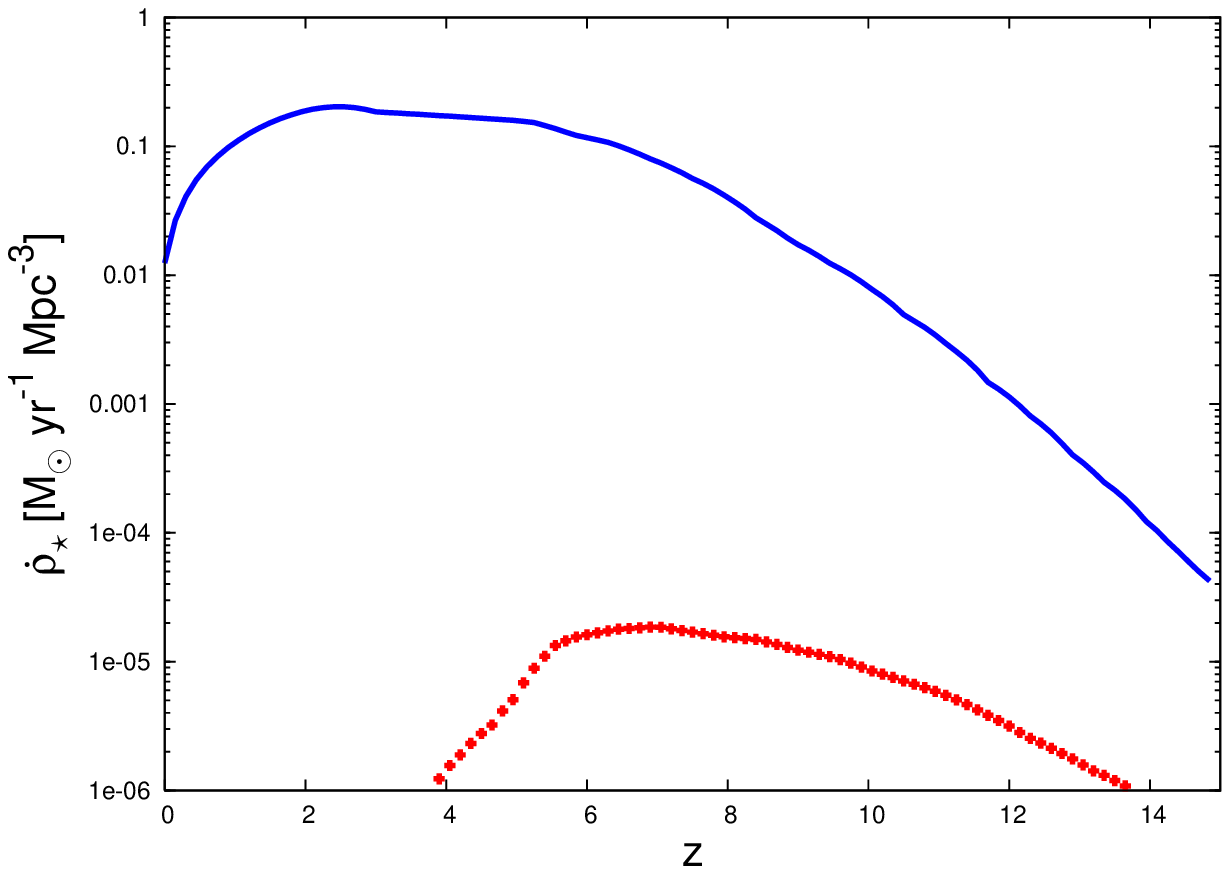}
\includegraphics[width=8.5cm,angle=360]{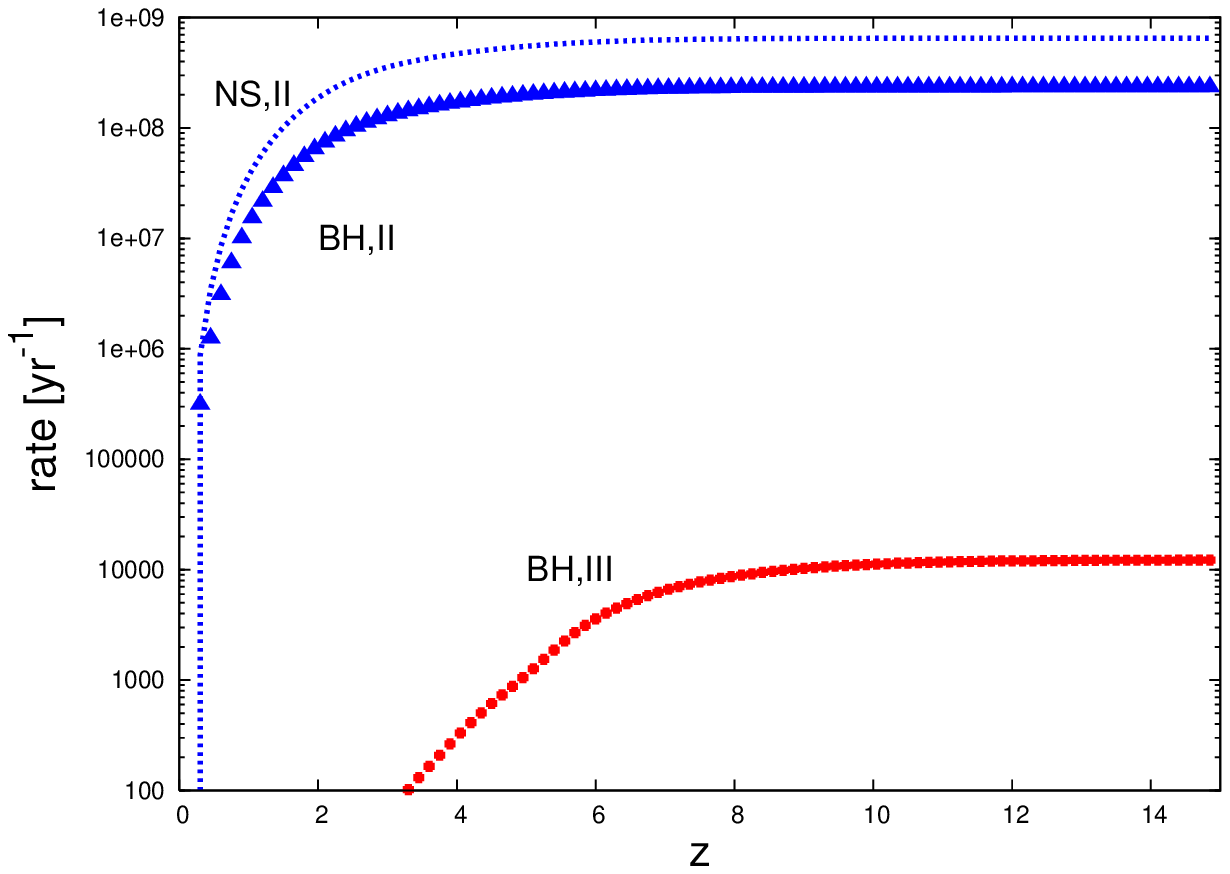}
\caption{{\it Top panel:} redshift evolution of the comoving star formation rate density (solid
lines). The dotted line shows the contribution of Pop~III stars (see text). {\it Bottom panel:}
redshift evolution of the number of gravitational wave sources formed per unit time within a
comoving volume. The top curves show the contribution of Pop~II stars leaving behind neutron
stars (NS,II solid line) and black holes (BH,II triangles); the bottom dotted line represents
the contribution of Pop~III stars collapsing to black holes (BH,III see text).}
\label{sim:sfr}
\end{figure}

The top panel in Fig.~\ref{sim:sfr} shows the redshift evolution of the cosmic star formation rate
and the contribution of Pop~III stars predicted by the simulation\footnote{The results
shown in Fig.~\ref{sim:sfr} refer to the fiducial run in \citet{TFS} with
a box of comoving size $L=10 h^{-1}$~Mpc and $N_{\rm p} = 2 \times 256^3$ (dark+baryonic)
particles.}. In this model, the critical metallicity which defines the Pop~III/Pop~II 
transition is taken to be $Z_{\rm cr} = 10^{-4} Z_{\sun}$, at the upper limit of the 
allowed range. However, additional runs show that decreasing the critical metallicity
to $10^{-6} Z_{\sun}$ reduces the Pop~III star formation rate by a factor $< 10$. 
It is clear from the figure that Pop~II stars always dominate the cosmic star formation
rate; however, in agreement with previous semi-analytic studies, the simulation shows
that Pop~III stars continue to form down to $z < 5$, although with a decreasing
rate. Over cosmic history, the fraction of baryons processed by Pop~III stars is predicted
to be $f_{b} = 2 \times 10^{-6}$. 

The number of gravitational wave sources formed per unit time out to a given redshift $z$ can
be computed integrating the cosmic star formation rate density, $\dot{\rho_\star}(z)$,
on the comoving volume element
and restricting the integral over the stellar IMF in the proper range of progenitor masses, that is 
\beq
&&R_{\rm NS,II}(z) =
\int_0^z dz' \frac{dV}{dz'}\frac{\dot{\rho}_{\star, \rm II}(z')}{(1+z')}
\int_{8  M_\odot}^{20 M_\odot} dM \Phi(M)\\
&&R_{\rm BH,II}(z) =
\int_0^z dz' \frac{dV}{dz'}\frac{\dot{\rho}_{\star, \rm II}(z')}{(1+z')}
\int_{20  M_\odot}^{100 M_\odot} dM \Phi(M)\\
&&R_{\rm BH,III}(z)=\int_0^z dz'\frac{dV}{dz'}
\frac{\dot{\rho}_{\star, \rm III}(z')}{(1+z')}\nonumber
\left( \int_{100 M_\odot}^{140 M_\odot}dM \Phi(M)\right.\\ 
&&+\left.\int_{260 M_{\odot}}^{500 M_{\odot}} dM \Phi(M)\right).
\eeq
\noindent
where the factor $(1+z)$ at the denominator takes into account the time-dilation effect and
the comoving volume element can be expressed as,
\beq
&&dV=4 \pi r^2  \left(\frac{c}{H_0}\right) \epsilon(z) dz \\
&&\epsilon(z) = \left[\Omega_M(1+z)^3+\Omega_\Lambda \right]^{-\frac{1}{2}}.\nonumber  
\label{comvol}
\eeq
\noindent
Because of the different gravitational wave emission mechanisms, we have explicitly divided 
Pop~II sources between those that lead to the formation of a neutron star remnant (NS,II) 
and those which collapse to black holes (BH,II). 
Pop~III sources have been considered to be all stars which collapse directly 
to black holes (BH,III). Therefore we cut the mass range $140M_{\sun} \leq M \leq 260M_{\sun}$ 
of pair instability supernovae because in this range of masses the progenitors encounter the 
electron-positron pair creation instability and are completely disrupted by the explosion 
\citep{HW}. Note, however, that these results strictly apply only to non-rotating
stellar models and that rotation may affect mass-loss, nucleosynthesis, and the pre-supernova
structure. Recent analysis show that, at least for metal-free stars ($Z=0$), up to 10\% of 
the initial stellar mass can be lost when the critical break-up limit is reached but this
is not important enough to change their fate \citep{EMCHM}.  
The results are shown in the bottom panel of Fig.~\ref{sim:sfr}. It is evident that
the rate of Pop~III stars collapse to black holes at $z \geq 6$ is $\sim 10^{-5}$ times smaller than the 
rate of gravitational wave sources associated to Pop~II stars. The latter are
dominated by core-collapse SNe leaving neutron star remnants, because of the adopted shape of
the stellar IMF. 

To compute the gravitational signal emitted by each single source, it is important to estimate
the fraction of the initial stellar mass which ends up in the final remnant NS or BH.
For Pop~III stars, we will neglect mass loss and assume that the final BHs have the same
mass of their progenitor stars ($100 - 500 M_{\sun}$). For Pop~II stars with masses in
the Type-II SNe progenitor range ($8 - 40 M_{\sun}$), the mass of the remnant is computed 
according to the original Woosley \& Weaver (1995) grid; for larger masses ($> 40 M_{\sun}$)
the mass of the final BH is taken to be equal to the helium core mass before the collapse,
$M_{\rm He}=\frac{13}{24}(M-20M_{\sun})$.

\section[]{Pop III stars as gravitational wave sources}
\label{sectionGWPopIII}

Recent studies have shown that the collapse of Pop~III stars
to black holes could be a much more efficient source of gravitational waves than
today's supernovae populations \citep{FWH,STKS1}. An estimate 
of the total energy released in gravitational waves during the first 
few seconds of collapse has been done in \citet{FWH}, 

\be 
\rm{E_{GW}\simeq 2\times10^{-3}M_{\sun}c^2}.
\label{EgwFryer}
\ee

\noindent
In their two-dimensional numerical simulations the authors have considered the entire life 
and death of a zero-metallicity star of $300M_{\sun}$, taking the effects of 
general relativity and neutrino transport\footnote{Neutrino transport
  is followed in a parameter-free way.} into account. As first noted 
in \citet{FWH}, the main features of the Pop~III collapse are
different from those that appear in ordinary core-collapse SNe. 
In the latter case, core collapse is halted when the central density
exceeds $\sim 10^{14}$g cm$^{-3}$ and when neutron degeneracy and
nuclear forces become relevant. Conversely, the core of a $300M_{\sun}$ star 
is so large that it collapses into a black hole before nuclear forces can 
affect it \citep{STKS1,FWH}. 

In addition, as noted by \citet{FWH} and \citet{STKS1}, rotation 
and higher central temperatures in Pop~III progenitors, may halt 
the collapse of the core and produce a {\itshape weak thermal bounce} 
at lower central density of a few times $\sim 10^{12}$g cm$^{-3}$. 
After this bounce, the final fate of hot neutronized core, 
as soon as entropy gradients becomes negligible, is to collapse to a black hole.

Lacking more detailed numerical models on the collapse of Pop~III stars,
\citet{BSRJM} have modelled the gravitational wave signal emitted in the collapse of a 
$300M_{\sun}$ Pop~III star using as a template the waveform of an ordinary core-collapse
supernova,

\be
f|\tilde{h}(f)|=\frac{G}{\pi c^4 r}E_{\nu}<q>\left[1+\frac{f}{a}\right]^3e^{-f/b},
\label{specshape}
\ee

\noindent
as obtained by \citet{MRBJS}
using sophisticated two-dimensional numerical simulations.  The 
stellar core-collapse is taken to be axially symmetric and is based on 
a detailed implementation of Boltzmann's equation for neutrino
transport (\citealt{MRBJS}). In equation (\ref{specshape}), $\tilde{h}$ is the 
Fourier transform of $h$, G is Newton's constant, $E_{\nu}$ is the total 
energy emitted in neutrinos, $<q>$ is the average value of the anisotropy 
parameter (defined in \citealt{MJ}), $r$ is the distance of the supernova 
event, $a$ and $b$ are respectively 200Hz and 300Hz. 
This choice of $a$ and $b$ values is motivated in order to reproduce the
spectral shape of the s15r model simulated in \citet{MRBJS}. 

The energy spectrum adopted by \citet{BSRJM}  therefore  is

\be
\frac{d\rm{E_{GW}}}{df}=\frac{16\pi^2c^3r^2}{15G}f^2|\tilde{h}(f)|^2\qquad \rm{[erg/Hz]},   
\label{singlespec} 
\ee

\noindent
with the additional requirement that the total energy emitted had the same 
value found by \citet{FWH}, 

\be
{\rm E_{GW}}=\frac{16\pi^2c^3r^2}{15G}\int dff^2|\tilde{h}(f)|^2=2\times10^{-3}M_{\sun}c^2.  
\ee

\begin{figure}
\includegraphics[width=6.0cm,angle=270]{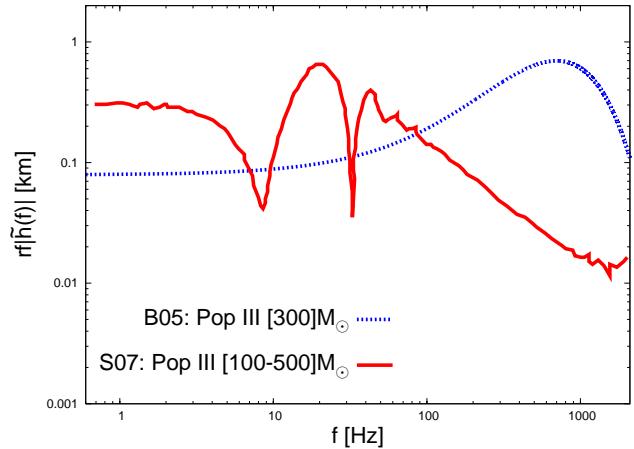}
\caption{Spectral distribution of GWs
  calculated in \citet{STKS1} (model S07, solid line) and in
  \citet{BSRJM} (model B05 with $E_\nu = 10^{55}$~erg and $<q>=0.03$, 
dotted line).}
\label{popIIIwaveforms}
\end{figure}
\noindent
The resulting GW spectrum is shown in Fig.~\ref{popIIIwaveforms}. 

More recently, \citet{STKS1} have studied the rotational collapse of Pop~III stars
in the same mass range. These studies have been performed using
two-dimensional hydrodynamic simulations, using a realistic equation 
of state based on the relativistic mean field theory \citep{Shen}, and
taking neutrino cooling and state-of-the-art reactions of neutrinos into account. 
For comparison with the previous study of \citet{FWH}, they discuss the GW emission from the
collapse of a $300M_{\sun}$ Pop~III star and assume the same initial core configuration.  
The total energy emitted in gravitational waves is found to be,
$\rm{E_{GW}}\sim 0.2\times 10^{-3}M_{\sun}c^2$, that is a factor 10 smaller than 
the result of \citet{FWH} (see eq.\ref{EgwFryer}), and the authors ascribe 
this discrepancy to the assumed different initial angular momentum distributions. 

In Fig.~\ref{popIIIwaveforms} the solid line shows the GW spectral 
distribution obtained in \citet{STKS1}. As it is clear from the 
comparison of the two curves, the contribution to GW emission 
from neutrinos in the model of \citet{STKS1} dominates the spectrum 
below 10 Hz and the matter contribution peaks at around 20 Hz,
whereas adopting the spectral energy distribution of ordinary supernovae,
as done by \citet{BSRJM}, these frequencies are shifted in the kHz range. 

As discussed by \citet{FWH} and \citet{STKS1}, these differences are expected, given 
the different nature of the progenitor and the distinct collapse dynamics. 
It is interesting to note, as shown by \citet{STKS2}, that since most of the
inner core is absorbed into the black hole, only about 
$10\%$ of the gravitational energy of the core can be carried away by
neutrinos (even in the case of the most rapidly rotating model). 
On the other hand, neutrinos carry away $99\%$ 
of the gravitational energy of proto-neutron stars in 
ordinary core-collapse supernovae. 

Unless otherwise specified, in our estimates of the gravitational wave
background produced by Pop~III stars, we will describe the single source 
energy spectrum using equation (\ref{singlespec}) with $f|\tilde{h}(f)|$  
taken from the results of \citet{STKS1}.

\section[]{Pop II stars as gravitational wave sources}
\label{sectionGWPopII}
In spite of the many elaborate studies carried out so far, the physics involved
in core-collapse supernovae is not yet completely understood. 
During the past years, theorists have done many efforts 
to give reliable predictions of the gravitational wave signals 
associated to these catastrophic events 
(for recent reviews we refer to \citealt{KST} and \citealt{Ott}).

In order to estimate the stochastic gravitational 
wave background from Pop~II stars, we have adopted  
different single source gravitational wave spectra 
obtained from the most sophisticated numerical 
simulations available in the literature.

The most recent calculations incorporate general relativity 
as well as the most relevant physics, such as a microphysical
finite-temperature nuclear equation of state, a scheme to
account for electron capture and neutrino 
losses during the collapse, an accurate treatment 
of the neutrino transport \citep{BLDOM,Ott2,Ott3,DOMJ2,DOMJ}. 

These studies seem to indicate that, in spite of the dependence of the 
gravitational wave burst on the precollapse central angular velocity, 
on the progenitor mass, on the equation of state, etc, all the 
simulated waveforms exhibit a generic bounce dynamics and signal morphology \citep{Ott2,Ott3,DOMJ2,DOMJ}. 
The bounce dynamics seems to be governed by the stiffening of the equation 
of state at nuclear density and, as a consequence, the resulting waveforms show one 
pronounced large spike at the bounce and a gradually damped ring-down \citep{DOMJ}. 
\subsection{Pop II progenitors with masses in the range [8-20]$M_{\sun}$}
To model the single source gravitational spectrum for Pop~II stars with masses in the range 
[8-20]$M_{\sun}$, we have used equation (\ref{singlespec}) adopting the $f|\tilde{h}(f)|$ predicted 
by the model labelled as s15r in \citet{MRBJS}. This choice provides the most optimistic conditions
for the total energy released in gravitational waves (\citealt{MRBJS}), and has been 
considered to be representative of the emission in core-collapse SNe
by previous studies (\citealt{BSRJM,SODV}). 
We have assumed that the total energy emitted in
gravitational waves is ${\rm E_{GW}} \simeq 1.8 \times
10^{-8}M_{\sun}c^2$ (that corresponds to a conversion efficiency 
of the mass energy in gravitational waves of $\epsilon \simeq
1.5\times 10^{-7}$, \citealt{MRBJS}).
The latest studies carried out by \cite{DOMJ2,DOMJ} confirm that this choice 
for the single source spectrum is representative of core-collapse supernovae with
progenitors in the above mentioned mass range.

Recently, \citet{BLDOM} have proposed an acoustic mechanism for core-collapse
supernovae associated to stellar progenitors in the mass range
$8-25M_{\sun}$. In this model, g modes (mainly l=1,2 core g
mode) are excited  since the proto-neutron star undergoes 
the standing-accretion-shock instability (SASI, see for a review on the
subject \citealt{BLDOM,Ott1,Ott}) induced by turbolence and accretion downstream of 
an unstable and deformed stalled supernova shock. This mechanism has been
suggested to 
be sufficient to drive a supernova explosion. 

In \citet{Ott1} the authors investigate this mechanism using a two dimensional 
axisymmetric newtonian simulation. The results seem to indicate 
that the emission process in core-collapse supernovae may be 
dominated by the oscillations of the proto-neutron star core.  
The authors explore different type of progenitors and find that the
more massive is the progenitors iron core, the higher are the
frequency and oscillation amplitudes and, as a consequence, the stronger
is the gravitational wave emission. It is interesting
to note that for these models the contribution 
to gravitational waves of anisotropic neutrino emission is completely 
negligible compared to that associated to core oscillations. 
In particular their estimate of the contribution of the anisotropic
neutrino emission (calculated as in \citealt{EPS,MJ}) is five times
smaller than previous one obtained in \citet{MRBJS}. 
The authors opinion is that this is due to the different approach
in treating neutrino transport. We refer the interested reader 
to the original paper for further details on the subject. 

The total energies emitted in gravitational waves are 
around ${\rm E_{GW}} \sim 10^{-8} M_{\sun} c^2$ (for a non rotating 
progenitor of 11 $M_{\sun}$ and a slowly rotating progenitor of 15
$M_{\sun}$) with the exception of a non rotating progenitor of 25
$M_{\sun}$, with a very massive iron core (1.92$M_{\sun}$), where 
the emitted energy is as large as ${\rm E_{GW}} \sim 8.2\times
10^{-5}M_{\sun} c^2$. It is important to note that these results 
are based on 2D Newtonian gravity simulations;  rotation and 3D 
simulations may change the structure of the SASI and consequently 
the energy emitted in gravitational waves, as point out by \citet{Ott1}. 
As an extreme and promising possibility, we also explore the gravitational background
produced by core-collapse SNe through this mechanism. 

\begin{figure}
\includegraphics[width=8.5cm,angle=360]{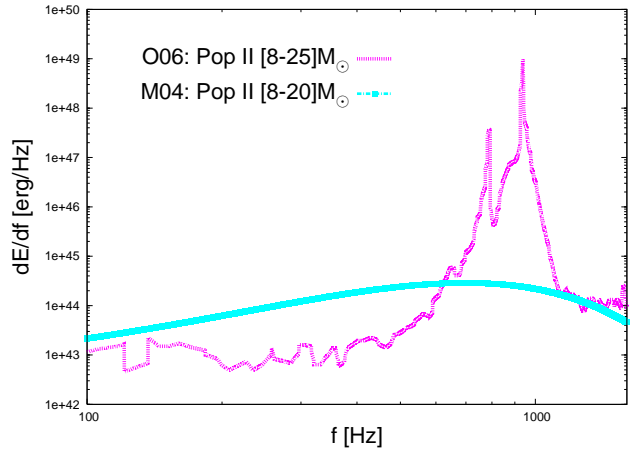}
\caption{The GW energy power spectra, for neutron star core-collapse,
  obtained from \citet{MRBJS} (model M04, with $E_{\nu} = 3 \times 10^{53}$~erg and $<q>=0.0045$) 
and \citet{Ott1} (model O06).}
\label{NSmodel}
\end{figure}

The energy spectra predicted by \citet{MRBJS} and \citet{Ott1} 
are shown in Fig.~\ref{NSmodel}. If we compare the two spectra, we see that the energy
spectrum obtained in \citet{Ott1} shows a pronounced
gravitational wave emission in the frequency range $600-1000$Hz, the
first burst of gravitational emission is centered about $800$Hz, and 
the second one near $950$Hz, due to the excitement of an l=2 core g-mode. 
\begin{figure}
\includegraphics[width=8.5cm,angle=360]{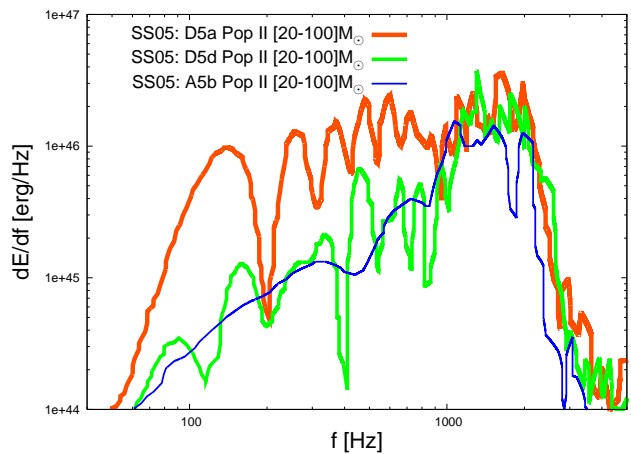}
\caption{The GW energy power spectra for models D5a, D5d and A5b obtained
in \citet{SS} (models SS05).}
\label{NSBHmodel}
\end{figure}

\subsection{Pop II progenitors with masses in the range [20-100]$M_{\sun}$}
Progenitors with masses between [20-100]$M_{\sun}$ may lead to a prompt  
collapse to black hole or to the formation of a proto-neutron star that,
due to the subsequent fallback will end its life as a black hole \citep{SS}.

Recently, \cite{SS} have run numerical simulations 
of the collapse of rotating massive cores 
associated to stellar progenitor masses in the range 
[51-98]$M_{\sun}$. These simulations have been performed 
in full general relativity using a parametric equation of state (\citealt{DFM1,DFM2})
that mimic a realistic one \footnote{This realistic equation of state
is the EOS Shen (\citealt{Shen}) and we note that this is the
same EOS used in \citet{Ott1} and \citet{DOMJ}.}. 
Gravitational waveforms are computed in terms of the quadrupole
formula taking into account only matter motions and neglecting
neutrino contribution (we refer the interested reader 
to the original paper for further details).

For the purpose of the present analysis, we adopt the gravitational wave
spectra from models D5a, D5d and A5b of \cite{SS}, which show 
the strongest gravitational wave emission with efficiencies ranging between
$0.2\times 10^{-6}\leq\epsilon\leq 0.3\times 10^{-6}$. Despite the fact that the
selected models are, in some cases, associated to a specific choice of the
progenitor stellar mass, we will neglect this dependence and assume that 
the same model holds for all progenitors in the $20 - 100 M_{\sun}$ range.
This approximation may be less accurate for black holes forming through
fallback from progenitors with masses $< 40 M_{\sun}$; in fact, the evolution
of the carbon/oxygen envelope as well as neutrino cooling (which will play a
role for longterm fallback with duration $>$ 100 ms) are not considered in
the simulations \citep{SS}.

In Fig.~\ref{NSBHmodel} we have plotted the energy power spectra for
models D5a, D5d, A5b. For all three models,  the 
spectra peak  at  $f \simeq 1-2$kHz. The energy 
power spectrum of model D5a is the result of a strong bipolar 
explosion (this feature is found for the first time in \citealt{SS}); 
it is broad in the low frequency region and flatter than 
those of the other two models. All spectra
peak at  $f \simeq 1-2$kHz. Due to the smaller 
progenitor mass ($51M_{\sun}$) the energy spectrum of model A5b 
shows a lower peak  then in model D5d
(progenitor mass $75M_{\sun}$) and shifted to lower frequencies.

\section{Gravitational Wave Background from Pop III/Pop II stars}\label{sectionGWB}

In this section we discuss the contribution of the different
gravitational wave sources to the background radiation. 
Following \citet{SFCFM}, the spectral energy density 
of the stochastic background can be written as 
\be
\frac{dE}{dS df dt}=\int^{z_f}_{0} \int^{M_f}_{M_i} dR(M,z)
 \big{<}\frac{dE}{dS df}\big{>}
\label{gwbk}
\ee
where $dR(M,z)$ is the differential source formation rate,
\be
dR(M,z)=\frac{\dot{\rho}_\star(z)}{(1+z)}\frac{dV}{dz}\Phi(M)dMdz,
\label{bkrate}
\ee
and $\big{<}\frac{dE}{dS df}\big{>}$, is the locally measured average 
energy flux emitted by a source at distance $r$. For 
sources at redshift $z$ it becomes,  
\be
\big{<}\frac{dE}{dS df}\big{>}=\frac{(1+z)^2}{4\pi d_L(z)^2}\frac{dE^{e}_{GW}}{df_e}[f(1+z)]
\label{singspec}
\ee
where $f=f_e(1+z)^{-1}$ is the redshifted emission frequency $f_e$, 
and $d_L(z)$ is the luminosity distance to the source.
It is customary to describe the GWB by a
dimensionless quantity, the closure energy density
 $\Omega_{\rm GW}(f) \equiv {\rho_{cr}}^{-1}(d\rho_{gw}/dlog f)$, 
which is related to the spectral energy density by the equation
\be
\Omega_{\rm GW}(f)=\frac{f}{c^3\rho_{cr}}\left[\frac{dE}{dS df dt}\right] ,
\ee
where $\rho_{cr}=3H_0^2/8\pi G$ is the cosmic critical density.

\begin{figure}
\includegraphics[width=8.5cm,angle=360]{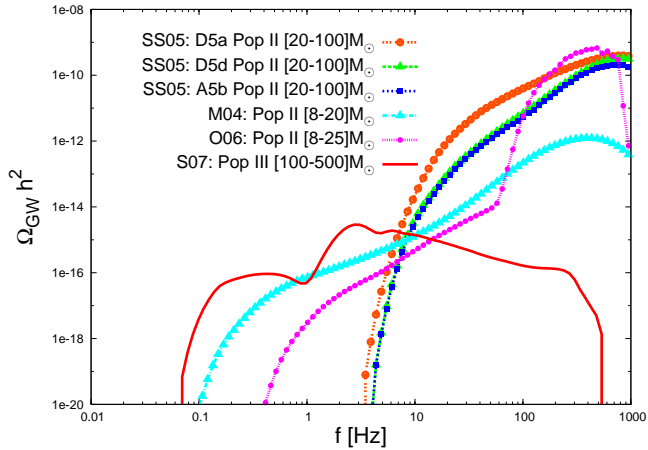}
\caption{The function $\Omega_{\rm GW}$ evaluated by assuming a frequency cut off
    $ f \leq$1 Hz.}
\label{allomega1}
\end{figure}

In Fig.~\ref{allomega1}, we show the function $\Omega_{\rm GW}$ evaluated by assuming 
a frequency cut-off $f \leq$1 Hz in the single source spectrum. 
This choice is motivated by the fact that current numerical 
simulations describe the source collapse at most for a few seconds, and 
cannot predict the emission below a fraction of Hz.
We see that below $\sim 10$Hz the Pop~III GWB dominates over
the Pop II background (except for a very small
region around $\sim1$Hz, where the Pop~III GWB has a minimum); its  maximum 
amplitude is $\Omega_{\rm GW}h^2\simeq 3\times 10^{-15}$ at $f=2.74$~Hz.
At larger frequencies the background produced by  Pop~II stars is 
much larger than that of  Pop~III.
Stars with progenitors in the range (20-100)$M_{\sun}$ contribute with  a nearly
monotonically increasing behavior, reaching amplitudes
$\Omega_{\rm GW}h^2\simeq 4\times 10^{-10}$ at $f\in (759-850) $ Hz.
For stars with progenitors in the range (8-25)$M_{\sun}$ the GWB
has a shape which depends on the waveform used as representative of the
collapse: that obtained using the waveform of \citet{MRBJS} 
is quite smooth and peaks at $f = 387$ Hz, with amplitude $\Omega_{\rm GW}h^2\simeq10^{-12}$. 
The background obtained using  waveforms produced by the  core-collapse model
of \citet{Ott1}, instead, is comparable in amplitude with those produced by more massive 
progenitors, reaching  the maximum amplitude $\Omega_{\rm GW}h^2\simeq 7\times 10^{-10}$ at $f= 485$ Hz. 

To evaluate the background in the low frequency region  
$f \leq$ 0.1 Hz, we can apply the so called 
{\it zero-frequency limit} (\citealt{Smarr,EPS,MJ,BSRJM}) and 
extend the single source waveform $f|\tilde{h}(f)|$ to lower 
frequencies (using equation \ref{specshape} for model M04, 
and tabulated values for models S07 and O06), where the emission 
is dominated by the neutrino signal (see Fig.~\ref{popIIIwaveforms}). 

\begin{figure}
\includegraphics[width=8.5cm,angle=360]{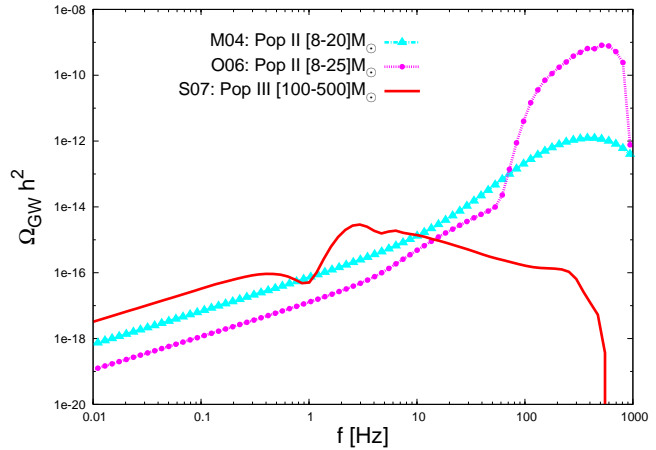}
\caption{The  function $\Omega_{\rm GW}$ in the zero-frequency limit.}
\label{allomega2}
\end{figure}
The results are shown in Fig.~\ref{allomega2}, where we plot
$\Omega_{\rm GW}h^2$ for all models shown in Fig.~\ref{allomega1},
with the exception of the Pop II signals derived from the \citet{SS}
single source spectra because in these computations neutrinos contribution 
is neglected and the extension would not be appropriate.

In the low frequency region, $\Omega_{\rm GW} \propto f$ because $f|\tilde{h}(f)|$
tends to a constant as shown in Fig.~\ref{popIIIwaveforms}.
In addition, the Pop~III-$\Omega_{\rm GW}$ dominates over the
Pop II, because  the GW emission increases 
with the stellar progenitor mass (the more massive is the progenitor, 
the larger is the neutrino luminosity).

We can compare our results with those recently obtained by 
\citet{STKS1}, \citet{SODV} and \citet{BSRJM}, who adopt different Pop~III
star formation rates and/or  different single source GW waveforms.
We find that our predicted Pop~III GWB is comparable to the 
lower limit reported by \citet{STKS1}; they use the same single source
waveform, but a different  star formation rate derived from \citet{SODV} 
and artificially shifted downward by a factor 7000. For this reason, they 
predict a baryon fraction in Pop~III stars of $~10^{-5}$, a value which
is much closer to our numerical result (see Section \ref{sectionsfr}).

Our Pop~III background is smaller than that found by  
\citet{SODV} and \citet{BSRJM}. In \citet{BSRJM}, the authors make a crude estimate 
of the Pop~III star formation rate, assuming that all Pop~III stars form
in a burst at $z=15$ with the same mass (300 $M_{\sun}$), and varying 
the fraction of baryons which go into Pop III stars in the range 
$5 \times 10^{-7} - 10^{-3}$. Comparing with their lower limit background, our
result shows a peak at lower frequency and with smaller
amplitude, and a larger low-frequency limit. These differences are naturally explained by
the adopted single source waveform (see Fig.~\ref{singlespec}). 
Finally, \citet{SODV} show a cumulative GWB from Pop~III and Pop~II stars 
for a cosmic star formation model where the baryon fraction in Pop~III stars
is $7 \times 10^{-2}$, much higher than that predicted by the numerical
simulation of \citet{TFS} (see Section \ref{sectionsfr}). 
It is not surprising, therefore, that their signal
associated to Pop~III stars is significantly higher than in our analysis;
furthermore, the peak amplitude is shifted to frequencies $~50-100$~Hz
due to the difference in the assumed single source waveform 
(see Fig.~\ref{singlespec}). The differences between the Pop~II GWBs 
are less pronounced, as the predicted Pop~II star formation rates are
comparable, and the adopted waveform for progenitors with masses 
$8 - 20 M_{\sun}$ is the same; still, due to the range of efficiency
assumed in the present analysis for the collapse of $20 - 100 M_{\sun}$
progenitors \citep{SS}, our Pop~II GWB is $\sim 2$ times smaller in amplitude 
than that predicted by \citet{SODV}.     

\subsection{Detectability}
In this section we discuss the possibility that Pop~II and Pop~III
backgrounds might act as a foreground, limiting the sensitivity 
of the detectors to the primordial GWB produced during 
the inflationary epoch \citep{KTHH,Seto,Seto2}.

\begin{figure}
\includegraphics[width=8.5cm,angle=360]{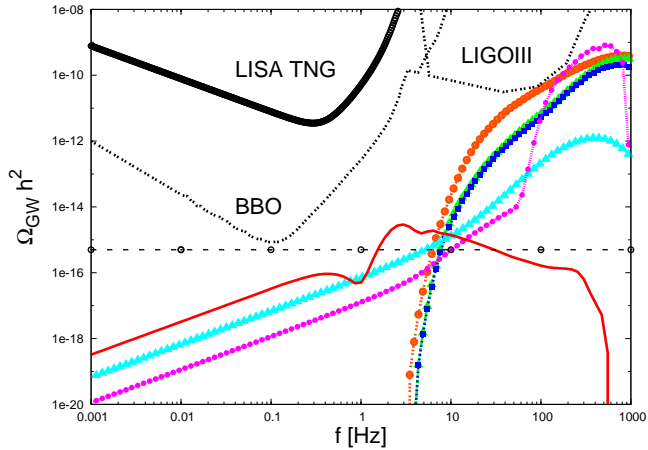}
\caption{The function $\Omega_{\rm GW}$ is plotted with the  sensitivity
  curves of space (LISA TNG, BBO Grand) and ground-based future detectors
  (LIGOIII).}
\label{omegavssens}
\end{figure}

\begin{figure}
\includegraphics[width=8.5cm,angle=360]{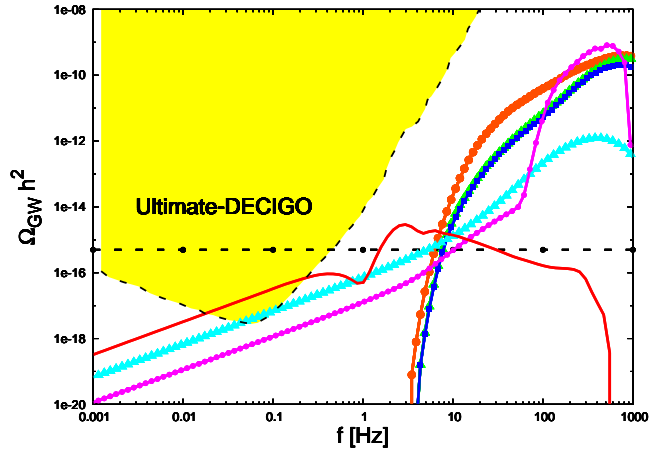}
\caption{The function $\Omega_{\rm GW}$ is plotted with the  sensitivity curve of
  Ultimate-DECIGO \citep{KTHH}. The horizontal dashed 
  line is the GWB spectrum produced during slow-roll inflation \citep{T}.} 
\label{omegavssens2}
\end{figure}

In Fig.~\ref{omegavssens} and Fig.~\ref{omegavssens2} we plot 
$\Omega_{\rm GW}$ as a function of frequency and 
the sensitivity curves of future detectors. 
In Fig.~\ref{omegavssens} we plot the sensitivity of 
LIGOIII used in \citet{B} and the detection limit for two proposed missions,
LISA TNG\footnote{http://www.srl.caltech.edu/~shane/sensitivity/} 
(The-Next-Generation) and BBO in 
the so-called BBO Grand configuration \citep{Seto}. In Fig.~\ref{omegavssens2} 
we plot the detection limit of DECIGO in the so-called Ultimate-DECIGO 
configuration, whose sensitivity is only limited by 
the standard quantum limit \citep{KTHH}.
All the sensitivity curves are obtained assuming correlated analysis 
of the outputs of independent spacecraft (or of different ground-based
detectors) \citep{B,KTHH,Seto}, with the exception of LISA TNG.
It is known that a single detector can only place upper limits on the
amplitude of a GWB.

The upper limit of the GWB generated in the Inflationary epoch, is given by the
dashed horizontal line in Fig.~\ref{omegavssens} and \ref{omegavssens2}.
It is derived from a recent update of cosmological
constraints on inflationary models (Cosmic Microwave Background 
measurements taken in combination with data from the Arcminute Cosmology
Bolometer \citet{KKMR1,KKMR2}). The data analysis
carried out by \citet{KKMR1,KKMR2} is
consistent with a primordial spectrum with no running spectral index
and zero tensor amplitude. The authors place an upper
limit to the tensor/scalar ratio $r < 0.35$ at  $95\%$  confidence level. 
According to these results, we have evaluated $\Omega_{\rm GW}$ from
equation (6) of \citet{T}, with $r=0.3$ and
no running of the spectral index of tensor perturbation.

As shown in Fig.~\ref{omegavssens},  LIGOIII, BBO and LISA TNG have no chance
to detect either the background produced by Pop~III/Pop~II stars and by
inflation.
From Fig.~\ref{omegavssens2} we see that the astrophysical backgrounds
are within the detection range of
Ultimate-DECIGO. However, they are smaller than the background generated in
the Inflationary epoch, which clearly dominates.
The Pop~III/Pop~II background becomes dominant only for $f\geq 3$ Hz.
It is worth reminding that the estimate of the Inflationary GWB shown in 
Fig.~\ref{omegavssens} and \ref{omegavssens2} is an upper limit.

\subsubsection{Duty Cycle}\label{Dutycycle}
An important parameter which quantifies the efficiency of the emitters
to generate a continuos signal in time domain is the duty cycle, $D$. 
This is defined as the ratio between the typical duration of the 
signal emitted by a single source, and the average time interval 
between two successive episodes of emission. 
When $D\rightarrow$1, the overall signal is continuous; conversely, 
if $D\ll1$, the resulting background is not continuous 
but rather characterized by a shot-noise structure.

An interesting feature of the GWBs produced by the 
collapse of Pop~III/Pop~II progenitors is that these 
backgrounds generate a signal with a peculiar 
shot-noise character, as suggested by \citet{FMS1,SFCFM}.   
In this case, the detector will receive a stationary 
sequence of bursts, with typical separations much 
longer than the average duration of each single burst.
The duty cycle measured by a local observer and generated by all
sources within the comoving volume out to a redshift $z$ can be written as
\be
D(z)= \int_0^z dR(z) \overline{\Delta\tau}_{gw}(1+z) 
\ee
where $\overline{\Delta\tau}_{gw}$ is the average time duration of 
individual signals and $dR(z)$ is the number of sources formed per unit
time at redshift $z$. 
\begin{table}
\caption{Duty Cycle for Pop~III/Pop~II progenitors.}
\begin{tabular}{|c|c|c|}
\hline
\multicolumn{3}{|c|}{$D$}\\
\hline
 NS(Pop~II) & BH(Pop~II)& BH(Pop~III)\\
\hline
 8.56$\times 10^{-2}$ & 3.10$\times 10^{-2}$&1.57$\times 10^{-4}$\\
\hline
\end{tabular}
\label{dutycycle}
\end{table}
We have calculated the duty cycle for Pop~II and Pop~III GWBs, assuming that 
the typical $\overline{\Delta\tau}_{gw}$ for these events is the dynamical timescale 
at core bounce, $\sim 1$ ms and $\sim 50$ ms respectively.
The results are illustrated in Table \ref{dutycycle} and show
that $D \sim 10^{-2}$ for Pop~II GWBs and $\sim 10^{-4}$ for
Pop~III GWBs.  

This shot noise structure of Pop~III/Pop~II GWBs 
might enhance the ability to discriminate these backgrounds from the dominant 
instrumental noise and from signals generated in the early Universe, even in 
frequency ranges where the latter have a higher amplitude.

\section{Gravitational Wave Background from Super Massive Stars}
\label{sectionSMS}

Recently, it has been suggested that Super Massive Black Holes (SMBH)
with masses of $\approx 10^5 - 10^6 M_{\sun}$ may form in protogalactic 
dark matter halos with virial temperatures $T_{\rm vir} \geq 10^4$~K
provided that (i) the gas has primordial composition and (ii) it is
irradiated by a strong ultraviolet (UV) flux \citep{BLO}. 
In these conditions, 3D hydrodynamics simulations show that, due to suppressed H$_2$
formation, the gas cools to temperatures that are only somewhat smaller than the virial
temperature of the host halo and condenses isothermally into very large clumps, with no
sign of fragmentation. As a result, the primordial cloud collapses into a 
central compact object containing $\geq 10$\% of the total baryonic mass, 
possibly through an intermediate phase of Super Massive Star (SMS) 
formation, (we refer the interested reader to \citealt{BLO} for further details). 
This scenario would provide potentially detectable sources of gravitational
waves and an alternative way to explain the presence of SMBHs with masses $\approx 10^9 M_{\sun}$
powering bright quasars at redshifts as high as $z \geq 6$ \citep{VR}.   
However, high redshift dark matter halos with $T_{\rm vir} \sim 10^4$~K are likely already enriched 
with at least trace amounts of metals and dust produced by prior star formation 
in their progenitors. Indeed, it has been shown that, even in the presence of a
sufficiently strong UV background, gas fragmentation is inevitable above 
a critical metallicity, whose value is between $Z_{\rm cr} \approx 3 \times 10^{-4} Z_{\sun}$ 
(in the absence of dust) and as low as $Z_{\rm cr} \approx 5 \times 10^{-6} Z_{\sun}$ for a 
dust-to-gas mass ratio of about $0.01 Z/Z_{\sun}$ (Omukai, Schneider \& Haiman 2008).  
Therefore, the actual rate of formation of (SMSs) SMBHs remains uncertain. 

Following \citet{BLO}, we can place an upper limit on the SMBH formation
rate at $z = 10$ requiring that their total mass density does not exceed the present-day
observed value (\citealt{MRGHMS,MH} and references therein), that is
\beq
&& f_{\rm smbh} \int^{\infty}_{M_{\rm min} (z=10)}
  \frac{dn(M,10)}{dM} M_{\rm smbh}(M)dM \\\nonumber
&& \leq 4.3 \times 10^5 (h/0.7)^2 M_{\odot} \hbox{Mpc}^{-3},
\label{eq:SMBHdens}
\eeq
\noindent
where $f_{\rm smbh}$ is the fraction of halos that contain a SMBH,  $M_{\rm smbh}$ 
is the SMBH mass which, following the results of the simulations, is assumed to
scale with the halo mass as $M_{\rm smbh} = 0.1 (\Omega_B/\Omega_M) M$,
$n(M,10)$ is the Press-Schechter halo mass function at redshift $z=10$, and $M_{\rm min}(z=10)$
is the dark matter halo mass corresponding to a virial temperature of $T_{\rm vir} = 10^4$~K
at $z = 10$, 
\be
M_{\rm min}(z=10) = 10^8 h^{-1} M_{\odot}  \left(\frac{\mu}{0.6} \right)^{-3/2} 
\left (\frac{\Omega_M^z}{\Delta_c} \right)^{1/2} \nonumber 
\ee  
\noindent
$\mu$ is the molecular weight, $\Delta_c=18 \pi^2 +82 d -39 d^2$, $d = \Omega_M^z -1$ and
\[
\Omega_M^z = \frac{\Omega_M (1+z)^3}{\Omega_M (1+z)^3 + \Omega_\Lambda} 
\]
for a flat cosmological model. For the adopted cosmological parameters, 
we find that $f_{\rm smbh} =  0.14$, that is about 14\% of protogalactic halos must verify the conditions suitable for
SMBH formation to reproduce the presently observed mass density. From eq.~(\ref{eq:SMBHdens})
we can also estimate the average SMBH mass, which is found to be $\sim 7.5 \times 10^6 M_{\sun}$.
\begin{figure}
\includegraphics[width=6.0cm,angle=270]{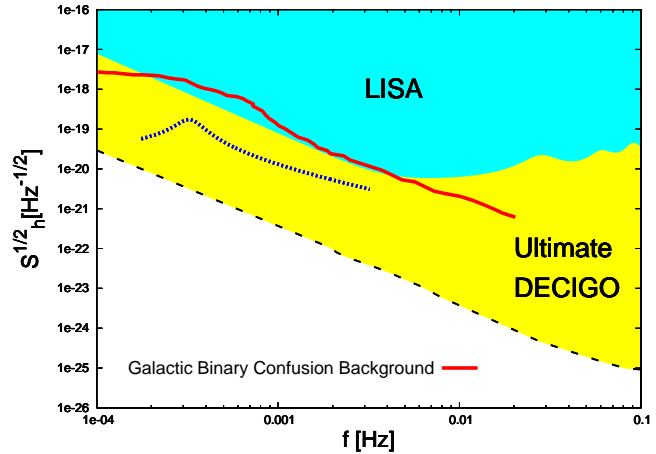}
\caption{The spectral strain amplitude of SMSs signal (dashed line) compared to the
  sensitivity curve of LISA and Ultimate-DECIGO. The Galactic binary confusion 
 background as reported in \citet{Schutz} is shown by the solid line.}
\label{SMSstrain}
\end{figure}
The gravitational collapse of a SMS to a SMBH has 
been studied by \citet{SBSS}, \citet{SS2} and \citet{LSS}.
In these papers, the authors suggest that the collapse of a SMS to
a SMBH may lead to the emission of an initial gravitational wave burst signal. 
In \citet{SS2} and \citet{LSS} the authors investigate 
the axisymmetric collapse of a uniformly rotating
SMS near the mass-shedding limit at the onset of radial instability. 
These simulations are performed in full general relativity using 
a polytropic equation of state (with $\Gamma =4/3$) and 
neglecting neutrino emission and transport (probably not relevant 
for very massive progenitors because of their low density and
temperature, as suggested by \citealt{LSS}). 
The authors estimate that $\sim 90\%$ of the 
total mass of the SMS is swallowed by the SMBH 
at the end of the collapse. Following \citet{SFCFM}, we 
have assumed that the single source spectrum emitted 
during the collapse of SMS can be modelled as a Lorenzian 
(for more details see also \citealt{SFCFM}),
\be
\frac{dE}{df}=\frac{\Delta \rm E_{GW}}{f_0
  N}\frac{f^2}{(f-f_0)^2+\Gamma^2}
\ee
where $f_0=c/10R_g$ with $R_g=2GM_{\rm smbh}/c^2$, 
$\Gamma=(2\pi\Delta t)^{-1}$, $\Delta t=1/f_0$ is the 
duration of the burst, $f_0N$ is the normalization and 
\be
\Delta \rm E_{GW}=\epsilon M_{\rm smbh} c^2
\ee
is the gravitational energy released in a burst, 
\be
M_{\rm smbh}= 0.1 (\Omega_B/\Omega_M) M_{\rm min}(z=10)
\ee
and $\epsilon$ is the efficiency. We assume that the SMS 
collapses to SMBH of comparable mass with an efficiency 
of $\epsilon=2.0\times 10^{-5}$. The adopted $\epsilon$ 
value was found by \citet{FWH} in the collapse scenario 
of very massive progenitors (see Section \ref{sectionGWPopIII}). 
This is a rough approximation of the gravitational wave spectrum 
but, in this uncertain evolutionary scenario, gives us the possibility 
to catch its main features. 
Using the estimated SMBH rate and the single source energy 
spectrum described above, we can compute the generated 
GWB using eq. (\ref{gwbk}). In Fig.~\ref{SMSstrain} we 
plot the spectral strain amplitude and the sensitivity 
curves of LISA and Ultimate-DECIGO. The estimated signal 
is too low to be detected by LISA and it lies above the foreseen 
sensitivity of Ultimate-DECIGO in the frequency range 
0.1~mHz$\leq f \leq$1~mHz; however, in this frequency interval
the sensitivity of Ultimate-DECIGO (as well as that of LISA) 
is limited by the unresolved background produced 
by the gravitational wave emission of Galactic compact binaries   
that acts as a confusion noise (\citealt{Nelemans}), shown in
Fig.~\ref{SMSstrain} with a solid line \citep{Schutz}. 
A possible future detection of this signal 
with Ultimate-DECIGO would require the application of 
sophisticated algorithms to data analysis, similar
to those that have been proposed for the LISA experiment (\citealt{CC}). 

\section{Discussion and conclusions}\label{sectionconclusion}

In this paper, we estimate the GWB produced by the
collapse of Pop~III/Pop~II progenitors. We use a new cosmic star 
formation history obtained from a recent numerical simulation
performed by \citet{TFS}. 
We find that our Pop~III/Pop~II GWB is below the 
sensitivity range of space detectors like BBO and LISA TNG 
in the present (proposed) configuration, but is in the 
sensitivity range of Ultimate-DECIGO, adding as a 
confusion-limited component. Differently to previous 
results of \citet{BSRJM}, \citet{SODV} and \citet{STKS1} 
we find that for $f\leq 2$ Hz, Pop~III GWB is masked by 
the GWB generated in the Inflationary epoch.

Clearly, the predicted amplitude of Pop~III/Pop~II GWB depends on 
the adopted star formation model and on the assumed single source 
GW spectrum, and we want briefly to point out the related
uncertainties. We have assumed, as a template for GW emission
associated to Pop~III stellar collapse, the waveform recently obtained 
by \citet{STKS1}.
The waveforms they find exhibit features that are
significantly  different  if compared to those found in 
the ordinary core-collapse SNe.  As noted by the authors,
their study is based on a Newtonian simulation, whereas a
detailed understanding of Pop~III stars collapse would
require a fully general relativistic approach \citep{STKS1}. 

In addition, the assumed Pop~II GW spectra are affected 
by the incomplete understanding of the SN explosion mechanism, 
by the complexity of the physics involved (equation of
state, structure of the stellar core, different approaches 
in treating neutrino transport, etc.) and 
by different approaches in numerical modeling. 

The variations among different single source gravitational 
wave spectra lead to comparable uncertainties in the corresponding
stochastic backgrounds: as an example, a two orders of magnitude difference
is found for the peak amplitude of the signal associated to Pop~II stars
leading to neutron star remnants, depending on whether the spectrum by \citet{MRBJS} 
or the upper limit provided by \citet{Ott1} is adopted.    

It is important to note that while the estimated backgrounds depend on the Pop~III star 
formation model, such as the adopted IMF and the value of $Z_{\rm cr}$, 
the advantage of our approach is that it allows a self-consistent 
description of the star formation and chemical evolution. 
Moreover, variations of the adopted $Z_{\rm cr}$ parameter 
within the allowed range ($10^{-6} - 10^{-4} Z_{\sun}$) 
lead to a factor $< 10$ decrease of the resulting 
Pop~III star formation rate. As long as Pop III stars
are assumed to be very massive, with $M > 100 M_{\sun}$,
the resulting star formation rate is not very sensitive to
the adopted IMF; this is because only stars with masses in the 
relatively small pair-instability progenitor range 
(140 - 260 $M_{\sun}$) contribute to metal enrichment. If
Pop III stars are assumed to form with masses $< 100 M_{\sun}$,
the smaller metal yields lead to a larger number of $Z < Z_{\rm cr}$
regions and therefore to a larger Pop III star formation rate.
Finally, additional 
simulations with different box sizes and resolution 
show that the scatter in the Pop~III/Pop~II ratio 
remains always $< 1\%$ \citep{TFS}. 

In addition, we have given an upper limit to the GWB produced by 
the collapse of SMSs to SMBHs, events that may occur in metal-free
halos with virial temperatures $T_{\rm vir} \geq 10^4$~K at $z = 10$ 
exposed to a strong UV background \citep{BLO}. 
LISA has no chance to detect 
this low signal and Ultimate-DECIGO, in this frequency region, 
is seriously limited by Galactic binary foregrounds. 

In conclusion, we want to point out that if future 
space missions, like Ultimate-DECIGO, will detect the Inflationary 
GWB, it will be necessary to disentangle this background from 
those produced by astrophysical sources.

Recent studies performed by \citet{DF} 
and \citet{Seto2} are moving in this direction suggesting 
different detection methods for non-Gaussian GWB. 
In particular \citet{Seto2} has proposed 
a method to check if the detected GWB is 
an inflation-type background or if it is contaminated 
by undetectable weak burst signals from Pop~III/Pop~II 
collapse. If, in the future, it will be possible 
to extract information on the Pop~III GWB 
component, we can have a tool to investigate 
the formation history of these first stars.   

\section*{Acknowledgments}
Stefania Marassi thanks the Italian Space Agency (ASI) for the support. 
This work is funded with the ASI CONTRACT I/016/07/0. 

\label{lastpage}

\end{document}